\documentclass[twocolumn,prb,superscriptaddress,showpacs]{revtex4-1}

\usepackage{amsmath,amssymb}
\usepackage{graphicx}
\usepackage{verbatim}
\usepackage{amsfonts}
\usepackage{dcolumn}
\usepackage{bm}
\usepackage{color}
\usepackage[colorlinks,citecolor=blue]{hyperref}
\usepackage{subfigure}

\begin{document}

\title{Non-Abelian $S=1$ Chiral Spin Liquid on the Kagome Lattice}

\author{Zheng-Xin Liu}
\affiliation{Department of Physics, Renmin University of China, Beijing 100872, China}
\affiliation{Institute for Advanced Study, Tsinghua University, Beijing 100084, China}

\author{Hong-Hao Tu}
\affiliation{Max-Planck-Institut f\"ur Quantenoptik, Hans-Kopfermann-Str.~1, D-85748 Garching, Germany}

\author{Ying-Hai Wu}
\affiliation{Max-Planck-Institut f\"ur Quantenoptik, Hans-Kopfermann-Str.~1, D-85748 Garching, Germany}

\author{Rong-Qiang He}
\affiliation{Department of Physics, Renmin University of China, Beijing 100872, China}
\affiliation{Institute for Advanced Study, Tsinghua University, Beijing 100084, China}

\author{Xiong-Jun Liu}
\affiliation{International Center for Quantum Materials and School of Physics, Peking University, Beijing 100871, China}
\affiliation{Collaborative Innovation Center of Quantum Matter, Beijing 100871, China}

\author{Yi Zhou}
\affiliation{Department of Physics, Zhejiang University, Hangzhou 310027, China}
\affiliation{CAS Center for Excellence in Topological Quantum Computation, University of Chinese Academy of Sciences, Beijing 100190, China}
\affiliation{Collaborative Innovation Center of Advanced Microstructures, Nanjing University, Nanjing 210093, China}

\author{Tai-Kai Ng}
\affiliation{Department of Physics, Hong Kong University of Science and Technology, Clear Water Bay Road, Kowloon, Hong Kong}

\begin{abstract}
We study $S=1$ spin liquid states on the kagome lattice constructed by Gutzwiller-projected $p_x+ip_y$ superconductors. We show that the obtained spin liquids are either non-Abelian or Abelian topological phases, depending on the topology of the fermionic mean-field state. By calculating the modular matrices $S$ and $T$, we confirm that projected topological superconductors are non-Abelian chiral spin liquid (NACSL). The chiral central charge and the spin Hall conductance we obtained agree very well with the $SO(3)_1$ (or, equivalently, $SU(2)_2$) field theory predictions. We propose a local Hamiltonian which may stabilize the NACSL. From a variational study we observe a topological phase transition from the NACSL to the $Z_2$ Abelian spin liquid.
\end{abstract}


\maketitle

\section{Introduction}

Topological order was used to describe and distinguish fractional quantum Hall (FQH) states \cite{TSG8259, L8395, W8987, WN9077, W9039}, and thereafter became a fundamental concept in condensed matter. In contrast to conventional long-range orders accompanied by spontaneous symmetry breaking in the Landau paradigm, topological orders are characterized by the topological degeneracy of ground states on a manifold with nonzero genus and fractionalized bulk excitations separated from the ground states by a gap. For instance, the FQH liquid at ${1\over3}$ filling has three degenerate ground states on a torus and its elementary excitations are particle-like `fractionalized' objects, such as the charge-${e\over3}$ quasi-holes. The charge-${e\over3}$ quasi-holes obey fractional statistics and are thus called anyons, i.e., the many-body wave function acquires a Berry phase $e^{i\pi/3}$ if one quasi-hole adiabatically exchanges its position with another one (this process is called braiding). More interestingly, the Pfaffian state proposed by Moore and Read \cite{MR9162} for the $\nu={5\over2}$ FQH liquid \cite{WES8776,RMM0899} supports non-Abelian anyons. After one braiding of two non-Abelian anyons, the degenerate many-body wave function undergoes a matrix `rotation' instead of a $U(1)$ phase-gate operation \cite{W9102, TQCreview}. Non-Abelian topological orders have potential applications in quantum information and quantum computation \cite{kitaev2003,TQCreview}.

Besides FQH systems, gapped spin liquids, such as resonating valence bond (RVB) states \cite{Anderson1987}, may also exhibit nontrivial topological orders. For example, the Kalmeyer-Laughlin chiral spin liquid \cite{KL8795} supports semionic anyons, and the short-range RVB state on a two-dimensional (2D) non-bipartite lattice carries $Z_2$ topological order \cite{moessner2001}. In seeking of spin liquids in realistic microscopic models, antiferromagnets on the kagome lattice have been widely studied \cite{Sachdev92, Ran07, Jiang08, white2011, schollwock2012,liao2017} due to its strong geometric frustrations, which is important for suppressing N\'eel order and favoring disordered ground states. On the experimental side, promising candidates of spin liquids have been synthesized, such as the Herbertsmithite realizing an $S=1/2$ kagome antiferromagnet \cite{THHan12}.  Recently, $S=1$ antiferromagnets have also attracted tremendous interest from experimental, theoretical and numerical sides. Several exotic $S=1$ spin liquid states, such as $U(1)$, $Z_2$, and (non-Abelian) chiral spin liquids, have been proposed \cite{greiter2009, LZN,bieri2012,wang2011,xu2012,yao2010,li2014,li2015, ZhouRMP17}.

In this paper, we construct both Abelian and non-Abelian $S=1$ spin liquid states on the kagome lattice. These wave functions are constructed by a Gutzwiller projection of $p_x+ip_y$-superconducting mean-field ground states within a fermionic slave particle representation of $S=1$ spins. It is shown that the topology of the mean-field states of the fermions determines the physical properties of the Gutzwiller-projected states. The projected topological, namely, weak-pairing, superconductors are $SO(3)_1$ [which is equivalent to $SU(2)_2$] non-Abelian chiral spin liquids (NACSL) \cite{greiter2009}. This is verified by comparing their modular matrices $T$ and $S$ with analytic results. We show that the NACSL exhibits a quantum spin Hall effect, where the spin Hall conductance is quantized to ${1\over2\pi}$. On the other hand, the projected trivial, namely, strong pairing, superconductors are Abelian $Z_2$ spin liquids without spin Hall effect. We propose a local Hamiltonian which may stabilize the NACSL as its ground state. By tuning the interaction parameters, we expect that there might be a topological quantum phase transition from the NACSL phase to the $Z_2$ Abelian spin liquid phase.

The rest of the paper is organized as follows. In Sec.~\ref{sec2}, we review the fermionic spinon representation of $S=1$ spins and describe the wave functions constructed from Gutzwiller projection of $p_x+ip_y$-superconductors. In Sec.~\ref{sec3}, we characterize the $SO(3)_1$ chiral topological order of the Gutzwiller-projected topological superconductor. In Sec.~\ref{sec4}, we propose a local Hamiltonian which may stabilize the NACSL. Based on a variational study, we suggest a possible topological quantum phase transition from the NACSL to a $Z_2$ Abelian spin liquid. Finally, Sec.~\ref{sec5} is devoted to a summary of our results.

\section{Construction of Gutzwiller-projected wave functions}
\label{sec2}

Let us start by introducing three species of fermionic slave particles (also called spinons)~\cite{LZN2,tu2013} $C_j=(c_{1j}, c_{0j}, c_{-1j})^T$ to represent the $S=1$ operators as
\begin{eqnarray*}
&&S^+_j=S^x_j+iS^y_j=\sqrt{2}(c_{1j}^\dag c_{0j}+c_{0j}^\dag c_{-1j}),\\
&&S^z_j=c_{1j}^\dag c_{1j}-c_{-1j}^\dag c_{-1j},
\end{eqnarray*}
under the local particle number constraint
\begin{eqnarray}\label{Constr}
c_{1j}^\dag c_{1j} +c_{0j}^\dag c_{0j}+c_{-1j}^\dag c_{-1j}=1.
\end{eqnarray}
In this way, the spin-spin interactions can be replaced by interactions between the fermions. The spin operators and the particle number constraint are invariant under the local $U(1)$ transformation $C_j\to C_je^{i\varphi_j}$, indicating a $U(1)$ gauge structure of this fermionic representation.

\begin{figure}[t]
   \centering
   \subfigure[Kagome lattice]{\includegraphics[width=0.23\textwidth]{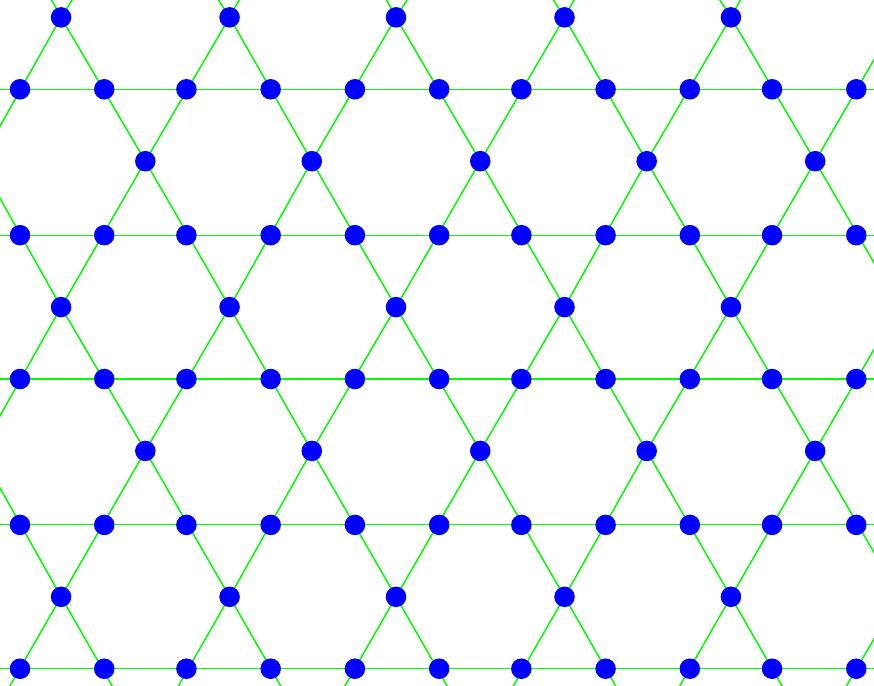}\label{fig_Kagome}}
   \hspace{0.1in}
   \subfigure[Spin pump]{\includegraphics[width=0.16\textwidth]{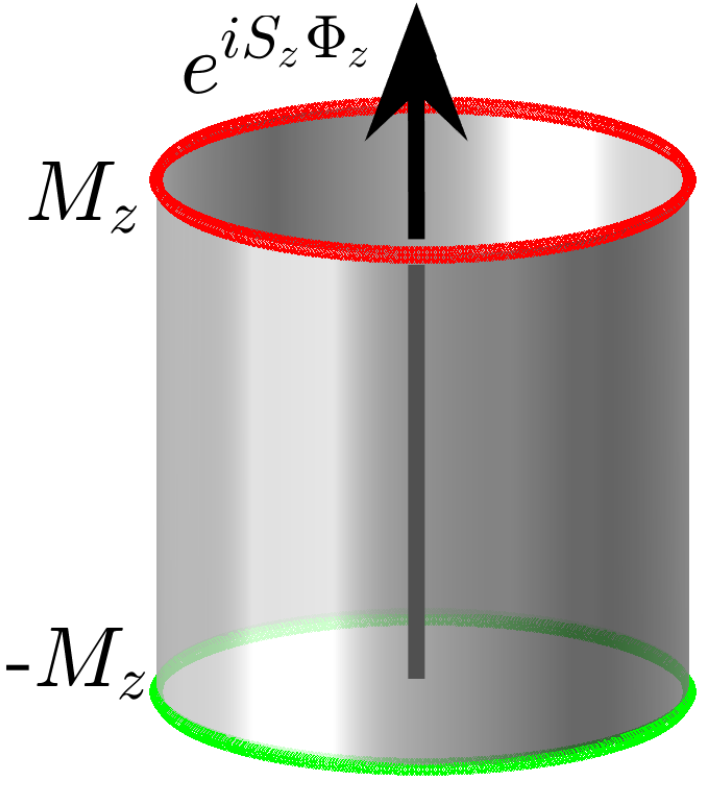}\label{fig_Pump}}
   \subfigure[Projected weak pairing state]{\includegraphics[width=1.6in]{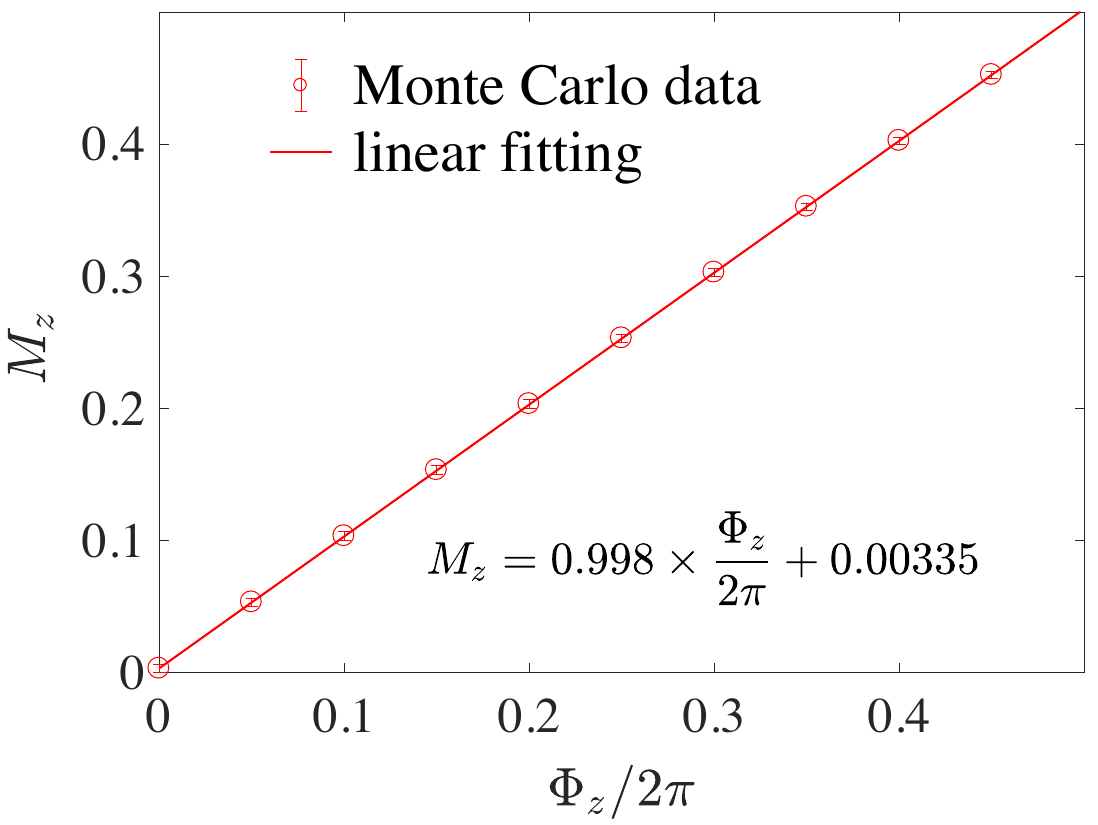}\label{fig_Pump_top}}
   \subfigure[Projected strong pairing state]{\includegraphics[width=1.6in]{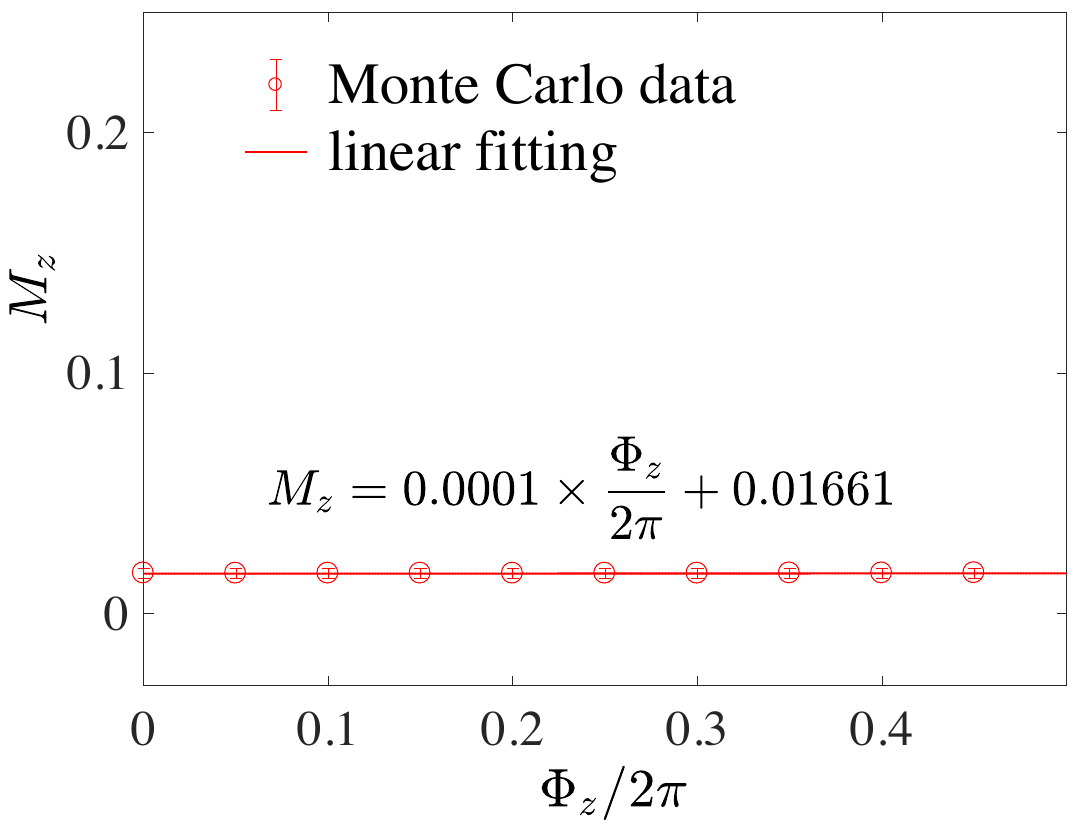}\label{fig_Pump_trv}}
   \caption{Laughlin's gauge invariant argument of quantum spin Hall effect. The calculation is performed on a cylinder with $10\times10$ unit cells (300 sites). (a) the structure of kagome lattice; (b) the pumping process; (c) projected topological superconductor (with $\chi=-1, \Delta=1, \lambda=1$) has quantized spin Hall conductance ${1\over2\pi}$; (d) projected trivial superconductor (with $\chi=-1, \Delta=1, \lambda=10$) has no spin Hall effect.}
   \label{fig:QSHE}
\end{figure}

We will focus on kagome lattice model with $SO(3)$ spin rotation symmetry [see Fig.~\ref{fig_Kagome}] that may harbor $S=1$ spin liquid phases. At the mean-field level, we consider the following quadratic Hamiltonian of fermions:
\begin{eqnarray}\label{MF}
H_{\rm mf}= \sum_{\langle ij\rangle} [\chi_{ij} C_i^\dag C_j + \Delta_{ij}C_i^\dag \bar C_j + {\rm h.c.}]
+ \sum_i \lambda_iC_i^\dag C_i,\nonumber\\
\end{eqnarray}
where $\bar C_j = (c_{-1j}^\dag , -c_{0j}^\dag, c_{1j}^\dag)^T$ which behaves in the same way as $C_i$ does under $SO(3)$ rotation since $C_i^\dag \bar C_j|{\rm vac}\rangle$ is a spin singlet ($|{\rm vac}\rangle$ is the vacuum state). The $\chi_{ij}$ term stands for spinon hopping and $\Delta_{ij}$ term represents the spinon pairing, $\lambda_i$ is the Lagrangian multiplier for the particle number constraint.  For simplicity, we assume that the mean-field parameters are site-independent, namely, $\chi_{ij}=\chi_{ji}^*=\chi$, $\Delta_{ij} = -\Delta_{ji}=\Delta e^{i\theta_{ij}}$ (where the phase of the pairing $e^{i\theta_{ij}}$ is bond-dependent and also depends on the pairing symmetry), and $\lambda_i=\lambda$ which plays the role of `chemical potential'. The three independent parameters $\chi, \Delta, \lambda$ are treated as variational parameters in our later discussion.  The Hamiltonian (\ref{MF}) is generated from $SO(3)$ symmetric interactions, e.g., antiferromagnetic Heisenberg interactions, through mean-field approximations and it respects the full spin-rotation symmetry \cite{LZN, LZN2, LiuZhouTuWenNg2012, LiuZhouNgExt} (see Sec.~\ref{sec4} and Appendix~\ref{appsec2}).

We can linearly combine the spinon creation operators $c_1^\dag, \, c_0^\dag, \, c_{-1}^\dag$ into the Cartesian bases,
\begin{eqnarray}\label{vector}
c_x^\dag={1\over\sqrt2}(c_{-1}^\dag-c_1^\dag),\ c_y^\dag={i\over\sqrt2}(c_{-1}^\dag+c_1^\dag),\ c_z^\dag=c_0^\dag,
\end{eqnarray}
then the mean-field Hamiltonian (\ref{MF}) is decoupled into three identical superconducting Hamiltonians for each species, $c_x, \, c_y, \, c_z$, namely, $H_{\rm mf}=H_{\rm mf}^x+H^y_{\rm mf}+H^z_{\rm mf}$, where
\begin{eqnarray*}
H_{\rm mf}^x= \sum_{\langle ij\rangle} [\chi  c_{x,i}^\dag c_{x,j} - \Delta e^{i\theta_{ij}}c_{x,i}^\dag c_{x,j}^\dag + {\rm h.c.}]
+ \sum_i \lambda  c_{x,i}^\dag c_{x,i},
\end{eqnarray*}
and similarly for $H^y_{\rm mf}$ and $H^z_{\rm mf}$. The $SO(3)$ symmetry thus manifests itself. The ground state of above mean-field Hamiltonian is a superconductor --- the familiar Bardeen-Schrieffer-Cooper wave function $|{\rm BCS}\rangle$ (see Appendix \ref{appGutz}).

Since the pairing term in (\ref{MF}) has odd parity $C_i^\dag \bar C_j=-C_j^\dag \bar C_i$, the pairing symmetry on the kagome lattice can be either $p$-wave or $f$-wave. Here we will restrict ourselves to the $p_x+ip_y$-pairing states where $\Delta_{ij} =\Delta e^{i\theta_{ij}}$, $\theta_{ij}$ is the angle between $x$-axis and the bond $(ij)$. It is known that the $p_x+ip_y$-superconductors contain two phase, the weak pairing phase and strong pairing phase, depending on the topology of its ground state~\cite{ReadGreen2000}. In the weak pairing region $-2|\chi|<\lambda <4|\chi|$, the mean-field state is a topological superconductor, where {\it each species of the spions $c_x, \, c_y, \, c_z$ carries Chern number 1}. On the other hand, in the strong pairing region $\lambda<-2|\chi|$ or $\lambda>4|\chi|$, the Chern number vanishes and the mean-field state is a trivial superconductor.

The mean-field theory indicated that there exist two topologically distinct $S=1$ spin liquid phases. In path integral language, the mean-field state corresponds to a saddle point and to obtain a more accurate description of the spin system we need to consider the fluctuations of the mean-field parameters around their saddle point values. Since the amplitude fluctuations have a gap, the low-energy physics are captured by the local phase fluctuations which behaves as a $Z_2$ gauge field coupling to the spinons (since the spinons are superconducting, the phase fluctuations of $\chi_{ij}$ and $\Delta_{ij}$ are Higgsed into $Z_2$). After integrating out the spinon fields, the low-energy effective field theory of the spin system depends on the topology of the mean-field state, if the spinons are in the strong pairing state the result is a $Z_2$ gauge field theory, while if the spions are in the weak pairing state the result is an $SO(3)_1$ Chern-Simons gauge theory (for details see Appendix \ref{appCS}).

Alternatively, a practical way of obtaining essential information of the spin liquids is manually enforcing the mean-field ground state to satisfy the particle number constraint (\ref{Constr}). By performing Gutzwiller projection to the ground state $|{\rm BCS}\rangle$ of the mean-field Hamiltonian (\ref{MF}) \cite{LiuZhouTuWenNg2012, Liucomplete14}
\begin{eqnarray}\label{GutzwillerWF}
|{\rm RVB}\rangle &=& P_G |{\rm BCS}\rangle,
\end{eqnarray}
an $S=1$ RVB wave function is obtained (see Appendix \ref{appGutz} for details of the projected state) according to Anderson's seminal result \cite{Anderson1987}. Here $P_G$ is the Gutzwiller projection operator ensuring that each site is singly-occupied by the fermions. We emphasize that, although before projection the spinon mean-field state may be a topological superconductor, it does not carry any intrinsic topological order, i.e., there is no topological degeneracy of ground states and no fractional bulk excitations. However, after Gutzwiller projection conditions are drastically changed. The resulting RVB wave function may carry Abelian or non-Abelian topological orders, depending on the strong pairing or weak pairing nature of the spinon mean-field states, as we will illustrate in the remaining part of this work.

In the strong pairing case, the Gutzwiller-projected trivial superconductor can be adiabatically connected to a nearest-neighbor RVB state, since the pairing amplitude of two $S=1$ objects in a ``Cooper pair" decays exponentially with their relative distance.   This ``short-range'' RVB phase has four-fold degenerate ground states on a torus and carries $Z_2$ Abelian topological order, similar to the $Z_2$ spin liquid phase for spin-1/2 systems. In the remaining part of this work, we will focus on the weak pairing case, the Gutzwiller-projected topological superconductor, which is more interesting.

\section{Identifying and Characterizing the non-Abelian chiral spin liquid}
\label{sec3}

In this section, we focus on the spin liquid phase corresponding to the topological superconductors. We first give a low-energy effective field theory of this spin liquid phase, and then numerically verify that various physical properties of the Gutzwiller-projected wave function are in well agreement with the field theory predictions. 

In the topological superconductor, each flavor of fermions carries a nontrivial Chern number, so we expect that the system has a nontrivial response when it is probed by $SO(3)$ symmetry twisting fields $A_{ij}$, where $A_{ij}=A^x_{ij}S^x+A^y_{ij}S^y+A^z_{ij}S^z$ behave like external $SO(3)$ gauge fields coupling to the spins. Because of the $SO(3)$ gauge invariance, we expect that, after integrating out the fermions and the $Z_2$ gauge fluctuations, the low-energy physics in the hydrodynamic limit is described by the following $SO(3)_1$ Chern-Simons theory
\begin{eqnarray*}
\mathcal L_{\rm res} = i{k\over 4\pi} {1\over2}{\rm Tr}[\varepsilon^{\mu\nu\lambda}(A_\mu\partial_\nu A_\lambda) - {1\over 3}A^3] + {\cal L}_{\rm Maxwell} + ...
\end{eqnarray*}
where repeated indices are summed over and $k=1$ since the Chern numbers of $c_x, c_y, c_z$ spinons are all equal to 1. As discussed in Appendix \ref{appCS}, the intrinsic field theory description of the system in the weak pairing case is still an $SO(3)_1$ gauge theory [see eq.~(\ref{intrin})], which predicts that the system falls in a NACSL phase. 

In the following, we provide evidence to illustrate that a Gutzwiller-projected topological superconductor is indeed a NACSL that is described by the $SO(3)_1$ Chern-Simons theory.

\subsection{Quantum spin Hall effect}\label{sec:SO3}

From the response field theory, if the probing field only contains a $z$-component, i.e. $A_\mu=A^z_\mu S^z$, as a response we obtain the spin Hall current
$$
J^z_\mu={\delta \mathcal L_{\rm res}\over \delta A^z_\mu} = {1\over 2\pi} F^z_{\mu},
$$
where $F^z_\mu= \sum_{\nu, \lambda} i\varepsilon^{\mu\nu\lambda}(\partial_\nu A^z_\lambda)$ is the strength of the probing field and the spin Hall conductance is quantized to 1 in unit of ${1\over 2\pi}$.

To verify above result, we study the response of the spin system using the Gutzwiller-projected wave functions. Based on Laughlin's gauge argument \cite{laughlin1981}, the spin Hall conductance can be obtained by measuring the spin pump of the Gutzwiller-projected state on a cylinder when a symmetry flux $\Phi_z=\oint \pmb A^z\cdot d\pmb x$ (see Fig.~\ref{fig_Pump}) is adiabatically inserted in the mean-field Hamiltonian 
\begin{eqnarray*}
H_{\rm mf}(\Phi_z) &=& \sum_{\langle ij \rangle} [\chi_{ij} C_i^\dag e^{iA^z_{ij}S^z}C_j + \Delta_{ij}C_i^\dag e^{iA^z_{ij}S^z} \bar C_j + {\rm h.c.}]
\\ &&+ \lambda \sum_i C_i^\dag C_i.
\end{eqnarray*} 
Accordingly, the Gutzwiller-projected ground state of above Hamiltonian is now a one-parameter family $|\mathrm{RVB}(\Phi_z)\rangle$ depending on the flux $\Phi_z$.

We compute numerically the total spin polarization $M_z$ (as a function of $\Phi_z$) accumulated in the vicinity of the upper boundary of the cylinder [the red circle in Fig.~\ref{fig_Pump}], which extends over several lattice sites depending on the width of the edge state. The spin polarization defined as 
$$M_z(\Phi_z)=\langle\mathrm{RVB}(\Phi_z)|S_z|\mathrm{RVB}(\Phi_z)\rangle$$ 
is a function of the flux $\Phi_z$, where $S_z=\sum_j S_j^z$ with $j$ running over the lattice sites covered by the gapless edge states. The numerical results for a weak pairing state (with $\chi=-1, \, \Delta=1, \, \lambda=1$) and a strong pairing state (with $\chi=-1, \, \Delta=1, \, \lambda=10$) are shown in Figs.~{\ref{fig_Pump_top}} and \ref{fig_Pump_trv}, respectively. It can be seen clearly that the projected topological superconductor has a spin Hall conductance $\frac{1}{2\pi}$, which agrees with the prediction of the $SO(3)_1$ Chern-Simons theory, while the projected trivial superconductor has no spin Hall conductance.

\subsection{Spin-spin correlation functions}

Similar to electronic quantum Hall states, the bulk of an $SO(3)_1$ NACSL should be gapped and the boundary is gapless.

To verify this expectation, we compute numerically the spin-spin correlation functions of the Gutzwiller-projected wave functions on a cylinder with $L_x\times L_y =20\times 10=200$ unit cells and periodic (open) boundary condition along the $x$ ($y$) direction. The variational parameters $\chi=-1, \, \Delta=1, \, \lambda=1$ fall in the weak pairing region. 

\begin{figure}[t]
   \centering
   \subfigure[ Correlation at the boundary.]{\includegraphics[width=0.4\textwidth]{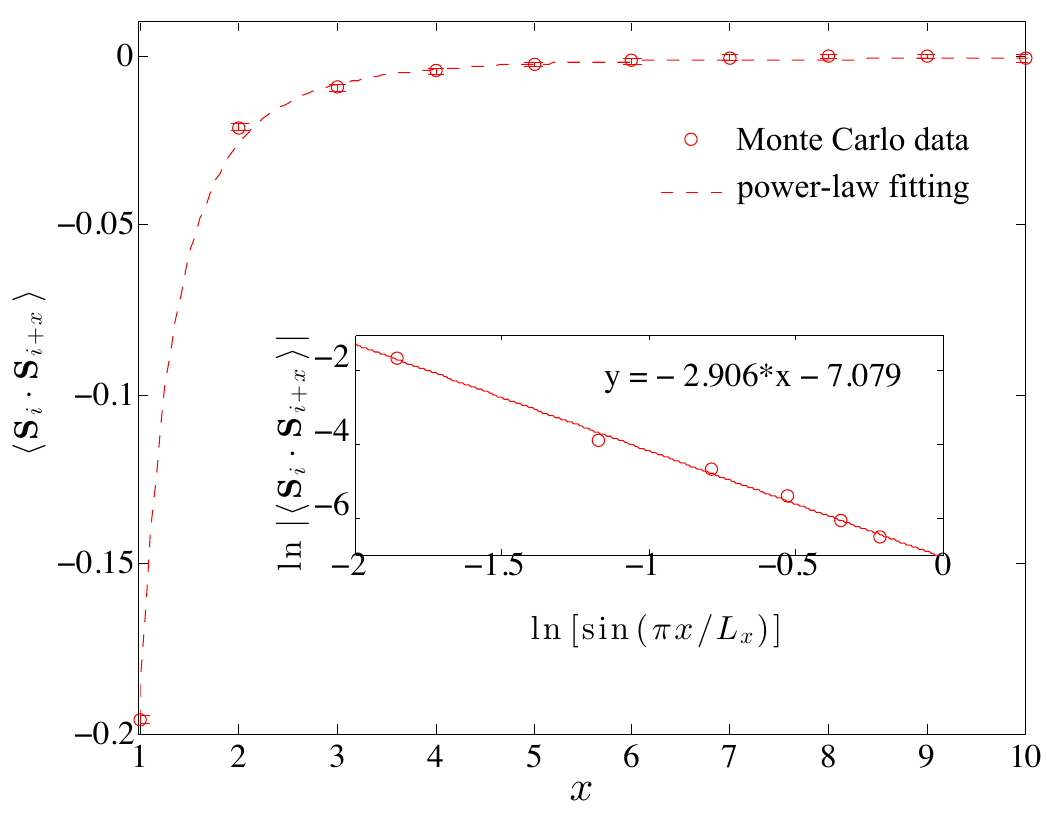}
   \label{fig_bdr} }
   \subfigure[ Correlation in the bulk.]{\includegraphics[width=0.4\textwidth]{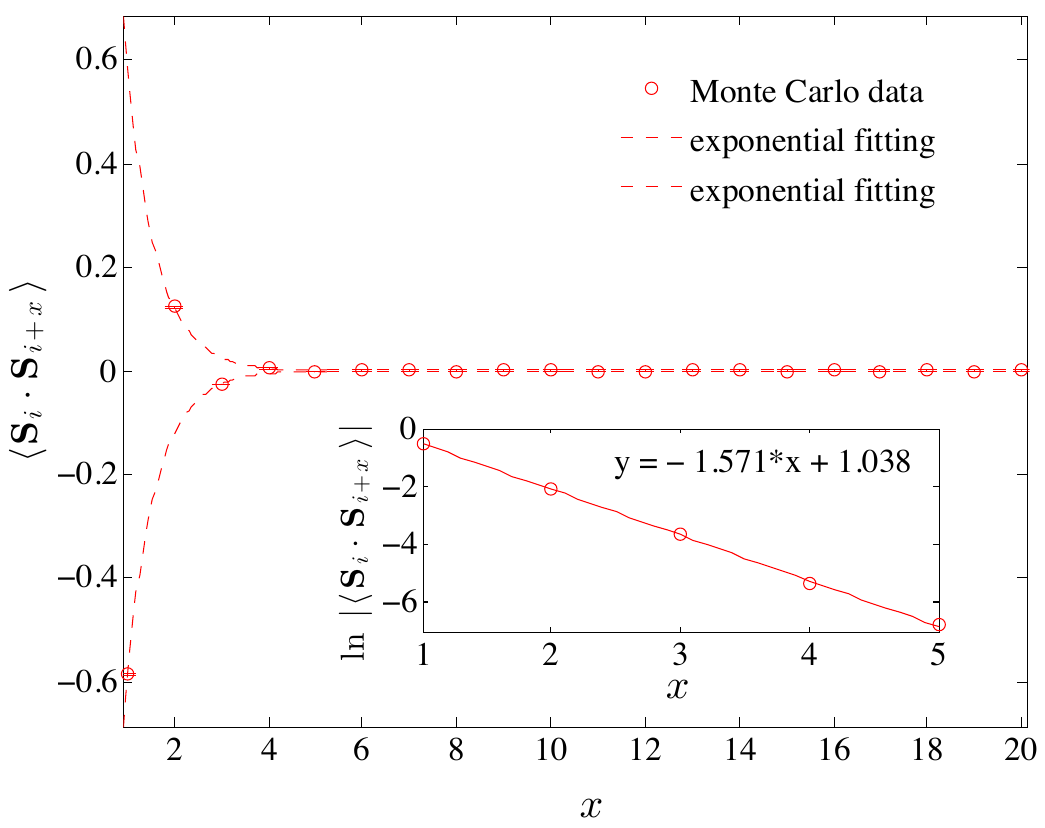}
   \label{fig_bulk} }
   \caption{The spin-spin correlations of the projected weak pairing state with parameters $\chi=-1, \Delta=1,\lambda=1$ are calculated on a cylinder with $20{\times}10$ unit cells. The $x$-direction has periodic boundary condition and the $y$-direction is open. (a) The spin-spin correlation function on the upper boundary has a power-law decaying behavior; (b) The spin-spin correlation function in the bulk (at the center of the cylinder) decays exponentially.}
   \label{fig:corr}
\end{figure}

The results in Fig.~\ref{fig:corr} demonstrate that the spin-spin correlation exhibits an exponential decay with distance in the bulk and a power-law decay with distance at the boundary, which implies that the bulk is gapped yet the boundary is gapless.

\subsection{Ground-state degeneracy}

The $SO(3)_1$ topological order (termed as $3^B_{3/2}$ in Ref.~\onlinecite{Wen15BTO}) has three types of anyons $I, \, \sigma, \, \psi$. Here $I$ is a trivial anyon for the vacuum, $\psi$ is a fermion, $\sigma$ is the Ising-like anyon and they obey the fusion rules $\sigma\times\sigma=I+\psi$, $\psi\times\psi=I$, $\sigma\times\psi=\sigma$. When the system is defined on a torus, there should be three degenerate ground states.

We now show that the three-fold degeneracy can indeed be reproduced by the Gutzwiller-projected wave functions. To this end, we note that inserting $Z_2$ gauge fluxes through the two non-contractible loops of the torus (equivalent to changing the boundary conditions for fermions from periodic to anti-periodic) does not change any local physical properties of the spin system, so there are four different mean-field states labeled by their $Z_2$ fluxes: $(0,0), \, (0,\pi), \, (\pi,0), \, (\pi,\pi)$.

However, in the weak pairing phase, not all of the four states survive after the Gutzwiller projection. Without loss of generality, we assume both $L_x$ and $L_y$ to be even. The mean-field state $(0,0)$ (with periodic boundary condition in both the $x$- and $y$- directions) vanishes after Gutzwiller projection because it has odd fermion parity. To see this, note that the pairing term is proportional to $k_x+ik_y$ at small momenta. For the $\Gamma$ point $\mathbf k=0$ in the weak pairing region $-2|\chi|<\lambda<4|\chi|$, fermions of all three species have vanishing pairing energy and negative kinetic energy, so the three modes $c_{x,\mathbf k=0}, \, c_{y,\mathbf k=0}, \, c_{z,\mathbf k=0}$ are occupied in the ground state. Away from the $\Gamma$ point, the modes $c_{\alpha,\mathbf k}$ and $c_{\alpha, -\mathbf k}$ form a Cooper pair so they are both occupied or unoccupied. Therefore, the fermion parity of the $(0,0)$ state is odd. In contrast, fermions cannot occupy the momentum $\mathbf k=0$ when the inserted fluxes are $(0,\pi)$, $(\pi,0)$ or $(\pi,\pi)$ because changing the boundary condition shifts the available momenta of fermions. All the fermions form Cooper pairs in the $(0,\pi)$, $(\pi,0)$, $(\pi,\pi)$ states so their fermion parities are even and they survive after Gutzwiller projection.

It should be further verified that the three states are orthogonal to each other and they form the three-dimensional ground-state subspace. To verify this, we consider a torus with $10{\times}10$ unit cells and the projected states for $\chi=-1, \, \Delta=1, \, \lambda=1.5$ are constructed. From Monte Carlo simulation with $3{\times}10^8$ steps, we find that $|\langle P_G(0,\pi) |P_G(\pi,\pi)\rangle| = 0.0012$, $|\langle P_G(\pi,0) |P_G(\pi,\pi)\rangle| = 0.001$, $|\langle P_G(\pi,0) |P_G(0, \pi)\rangle |= 0.0002$. This indicates that the three states are orthogonal to each other (up to errors of order $10^{-3}$) and we conclude that the degeneracy of the Gutzwiller-projected wave functions on the torus is indeed three.

\subsection{Modular $S$ and $T$ matrices}

The modular matrices $S$ and $T$\cite{WenST93, zhang2012}, as projective representation of the two modular transformations $\hat S=\left(\begin{matrix}0&-1\\1&0\end{matrix}\right)$ (a $90^\circ$ rotation) and $\hat T=\left(\begin{matrix}1&0\\1&1\end{matrix}\right)$ (a Dehn twist) on a torus (which generate a modular group), provide more information characterizing topological orders. The $SO(3)_1$ conformal field theory predicts that
\begin{eqnarray}\label{ST}
T = e^{-i\frac{\pi}{8}} \left(
\begin{matrix}
1 & 0 & 0 \\
0 & e^{i\frac{3\pi}{8}} & 0 \\
0 & 0 & -1
\end{matrix}\right),
S = \frac{1}{2} \left(
\begin{matrix}
1 & \sqrt{2} & 1 \\
\sqrt{2} & 0 & -{\sqrt{2}} \\
1 & -\sqrt{2} & 1
\end{matrix}
\right),
\end{eqnarray}
satisfying relations $S^2=(ST)^3=1$. The $T$ matrix is diagonal $T=e^{-{2\pi i\over 24}c_-}{\rm diag}(\theta_1,\theta_\sigma,\theta_\psi)$, where $\theta_i$ is the self-statistics of the $i$th anyon and is also called the topological spin. The $S$ matrix tells us the quantum dimensions of the anyons and their mutual statistics. For the $SO(3)_1$ theory, the quantum dimensions of the anyons $I, \, \sigma, \, \psi$ are $1, \, \sqrt2, \, 1$, respectively. The chiral central charge 
\begin{eqnarray}\label{cc}
c_-={3\over2}
\end{eqnarray}
can be obtained either from the diagonal $T$ matrix, or from the relation ${1\over D}\sum_i d_i^2\theta_i = e^{{2\pi i\over 8}c_-}$ \cite{Kitaev06AnnPhys}, where $D=\sqrt{\sum_id_i^2}=2$ is the total quantum dimension.

The modular matrices can be calculated from the projected states numerically, for instance, the $T$ matrix can be obtained from the universal wave function overlap
\begin{eqnarray}\label{DehnTwst}
\langle \psi_m|\hat T|\psi_n\rangle = e^{i\varphi_{m,n}},
\end{eqnarray}
where $\hat T$ is a Dehn twist of the torus, $|\psi_n\rangle$ are the degenerate ground states, and $\varphi_{m,n}$ are complex numbers in general (as illustrated below, we interpret $\hat T$ as an adiabatic process and $e^{i\varphi_{m,n}}$ as the Berry phases,  in which case $\varphi_{m,n}$ are real numbers). In our calculation, we choose the number of unit cells along the $x$- and $y$-directions as $L_x=L_y=L$. The quantity $\varphi_{m,n}$ has the scaling relation \cite{zhang2012, TuZhangQi13B,ZaletelMongPollmann13L,HeidarWen15L}
\begin{eqnarray}\label{alpha}
\varphi_{m,n}(L^2) = \alpha_{m,n} + \beta_{m,n} L^2 +o(L^2)
\end{eqnarray}
with the system size, where $\alpha_{m,n}$ is a universal signature of the topological order and $e^{i \alpha_{m,n}}$ is the $(m,n)$ entry of the modular $T$ matrix. It is numerically challenging to extract $\alpha_{m,n}$ because the wave function overlaps decrease exponentially as the lattice size increases. To overcome this difficulty, we apply the trick introduced in Ref.~\onlinecite{YouCheng} to divide the Dehn twist into many substeps such that in each substep the wave function changes adiabatically and the Berry phase can be obtained with a relatively high accuracy.

\begin{figure}[t]
\centering
\includegraphics[width=3.in]{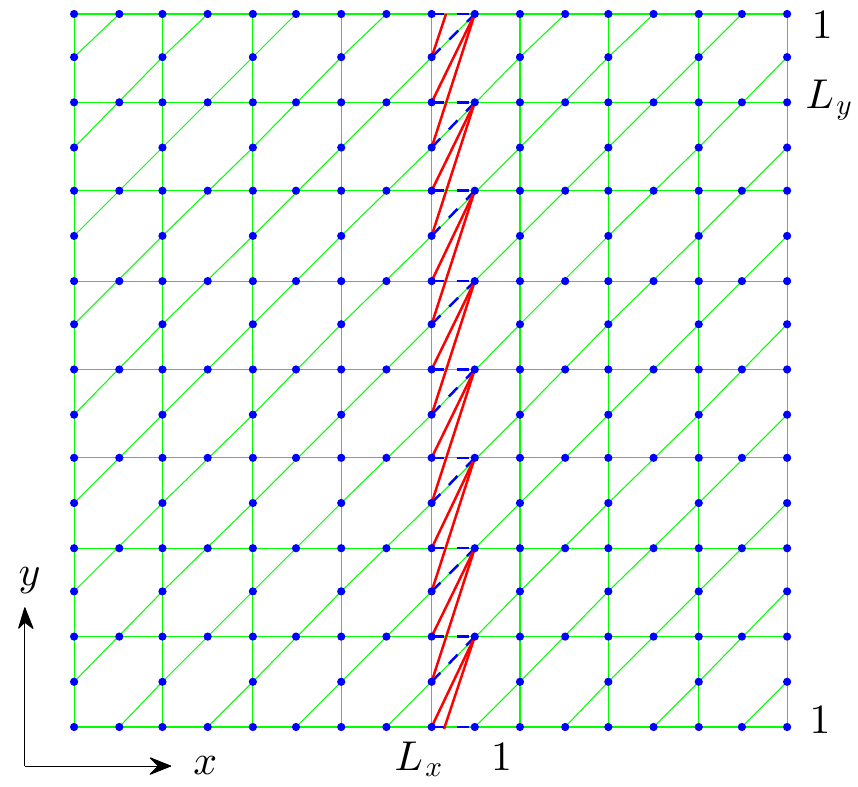}
\caption{(Color online) One step of the Dehn twist $\hat{T}_y$. The kagome lattice with $L_x = L_y$ is deformed to a square. The blue dashed lines represent the $x$-boundary couplings before the twist and the red solid lines represent the $x$-boundary couplings after one step of the Dehn twist. Some of the red lines across the $y$-boundary and the signs of the associated couplings are affected by the $y$-boundary condition. After a full Dehn twist, the $x$-boundary condition will be changed by the $y$-boundary condition.}
\label{fig:Ty}
\end{figure}

The Dehn twist can be performed either along $y$-direction with the operator $\hat T_y = \hat T$, or along $x$-direction with the operator $\hat T_x=\left(\begin{matrix}1&1\\0&1\end{matrix}\right)$. The two Dehn twists are not independent since $\hat{T}_x$ can be transformed into $\hat{T}_{-y}$ (the inverse of $\hat{T}_y$) by a global $90^\circ$ rotation. Therefore, we can only focus on the Dehn twist $\hat T=\hat T_y$. In calculating the modular matrix $T$, we first separate $\hat T_y$ into $L_y$ steps. In each step, the couplings across the $x$-boundary are shifted along $y$-direction by one lattice site (see Fig.~\ref{fig:Ty}) such that the Hamiltonian $H[(L_x,i),(1,i+t)]$ is transformed into the Hamiltonian $H[(L_x,i),(1,i+t+1)]$, where $H[(L_x,a),(1,b)]$ means that the unit cell $(L_x,a)$ at the $x$-boundary is `linked' with the unit cell $(1,b)$. In this process, some `links' cross the $y$-boundary and the signs of these `links' are affected by the $y$-boundary condition. After a full Dehn twist $T_y$, the boundary conditions $(0,\pi)$ and $(\pi,\pi)$ are shifted to $(\pi,\pi)$ and $(0,\pi)$, respectively, and the boundary condition $(\pi,0)$ remains unchanged. To make the twist more smooth, each step is further divided into several sub-steps by a parameter $\eta\in [0,1]$:
\begin{eqnarray}\label{substep}
H(t,\eta) &=& (1-\eta)H[(L_x,i),(1,i+t)] \nonumber \\
&&+ \eta H[(L_x,i),(1,i+t+1)],
\end{eqnarray}
which assumes the discrete values $\eta= 0, 0.25, 0.5, 0.75, 1$ in our calculations.

The Berry phase is extracted from the overlap of the projected ground states of above Hamiltonians
\[
\phi(t,\eta) = {\rm{Im}} \left[\ln \left(\langle t,\eta |P_G^\dag P_G|t,\eta+\delta\eta\rangle\right)\right],
\]
where $|t,\eta\rangle$ is the ground state of the mean-field Hamiltonian $H(t,\eta)$ and the total Berry phase is
\[
\varphi(L^2) = \sum_{t,\eta} \phi(t,\eta).
\]
For the sector $(\pi,0)$, the Hamiltonian goes back to itself after a full Dehn twist $\hat{T}_y$, so the evolution path is closed and the Berry phase is well defined. However, a full Dehn twist exchanges the two sectors $(\pi,\pi)$ and $(0,\pi)$ and the path is not closed, but we can perform the full Dehn twist twice to close the path and fix the Berry phase. The Berry phase for half of the loop (one full Dehn twist) is defined as half of that of a double twist. 

To simplify notations, we denote the three ground states $|P_G(\pi,0)\rangle$, $|P_G(0,\pi)\rangle$ and $|P_G(\pi,\pi)\rangle$ as $|x\rangle$, $|y\rangle$ and $|xy\rangle$, respectively. Then the two independent Berry phases in the Dehn twist $\hat T_y$ can be noted as $\varphi_{x,x}$ and $\varphi_{y,xy}$, which  respectively stands for the Berry phase of $|P_G(\pi,0)\rangle$ and $|P_G(0,\pi)\rangle$ (or $|P_G(\pi,\pi)\rangle$) gained in the adiabatic Dehn twist process. Similarly, the two independent Berry phases in $\hat T_x$ are $\varphi_{y,y}$ and $\varphi_{x,xy}$. As mentioned before, the Dehn twist $\hat{T}_x$ can be transformed into $\hat{T}_{-y}$ by a global $90^\circ$ rotation, therefore $\hat T_x$ and $\hat T_{-y}$ have the same Berry phase, which is contrary to the Berry phase of $\hat T_y$, so we have
\[
\varphi_{y,y} = -\varphi_{x,x}, \ \ \varphi_{x,xy} = -\varphi_{y,xy}.
\]

\begin{figure}[t]
\centering
\includegraphics[width=3.in]{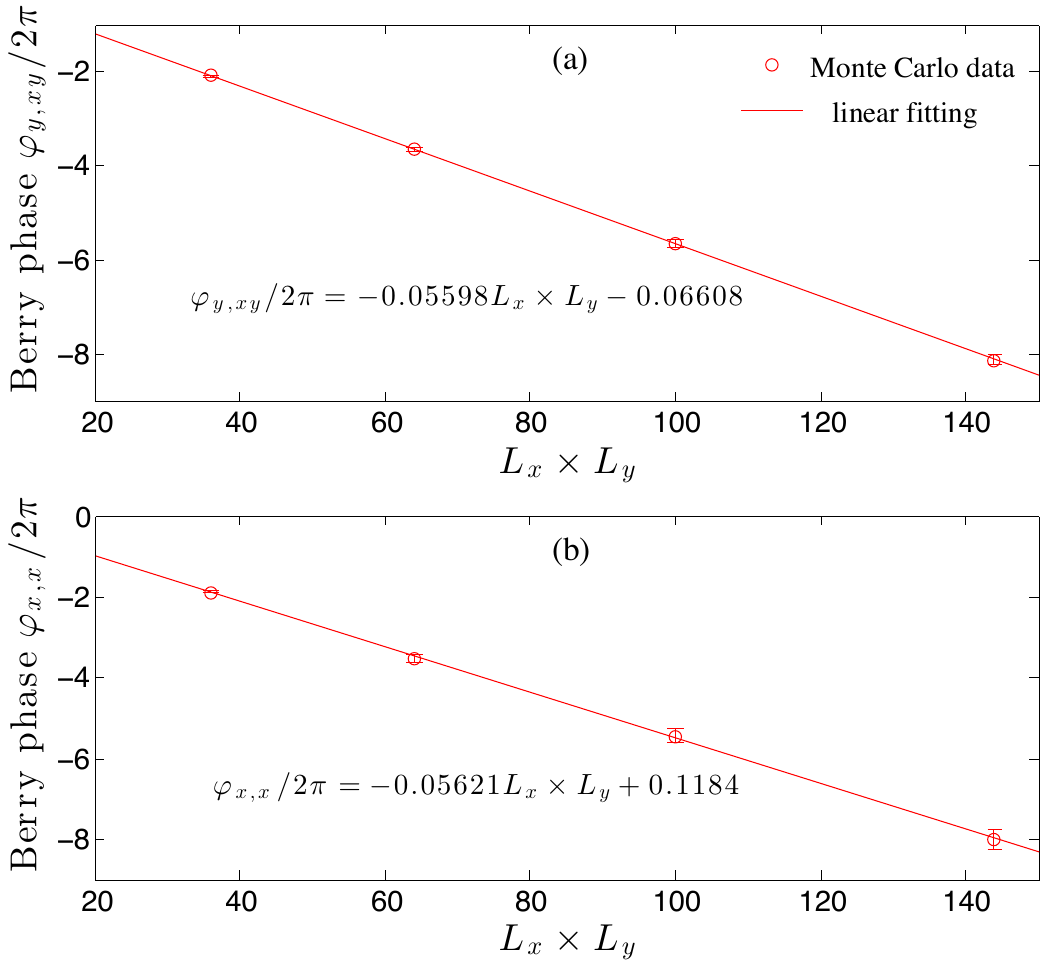}
\caption{(Color online) Scaling of the Berry phases (a) $\varphi_{y,xy}$ and (b) $\varphi_{x,x}$ for the Dehn twist $\hat{T}_y$ versus the system size. The system has $L_{x}$ and $L_{y}$ unit cells along the two directions and they are chosen to be equal for this calculation. The subscript $x,\, y$ and $xy$ in $\varphi$ denotes the three degenerate ground states $|P_G(\pi,0)\rangle$, $|P_G(0,\pi)\rangle$ and $|P_G(\pi,\pi)\rangle$, respectively.}
\label{fig:Ty_MM_MP}
\end{figure}

Figure \ref{fig:Ty_MM_MP} shows our Monte Carlo results of the scaling of the Berry phases $\varphi(L^2)$. Then from Eqs.~(\ref{DehnTwst}) and (\ref{alpha}) we obtain the modular matrix $T_y$
\begin{eqnarray}
\left(
\begin{matrix}
e^{ i\alpha_{x,x}} & 0 & 0 \\
0 & 0 & e^{ i\alpha_{y,xy}} \\
0 & e^{ i\alpha_{y,xy}} & 0
\end{matrix}
\right),
\end{eqnarray}
where $\alpha_{x,x} =0.2368\pi $ and $\alpha_{y,xy} =\alpha_{xy,y} = -0.1321\pi $ (the overlap between different topological sectors is omitted). Given the equivalence between $x$- and $y$- axes, we conclude that $T_x$ is
\begin{eqnarray}
\left(
\begin{matrix}
0 & e^{i\alpha_{x,xy}} & 0 \\
e^{i\alpha_{x,xy}} & 0 & 0 \\
0 & 0 & e^{i\alpha_{y,y}}
\end{matrix}
\right),
\end{eqnarray}
where $\alpha_{x,xy}=-\alpha_{y,xy}$ and $\alpha_{y,y}=-\alpha_{x,x}$ (they have been verified numerically). The modular $S$ matrix can be computed via $S=T_yT_x^{-1}T_y$.

Using a unitary transformation, the matrices $T_y$ and $S$ can be transformed into the standard form
\begin{eqnarray}
T=T_y = e^{i\alpha_{y,xy}} \left(
\begin{matrix}
1 & 0 & 0 \\
0 & e^{i(\alpha_{x,x}-\alpha_{y,xy})} & 0 \\
0 & 0 & -1
\end{matrix}\right)
\end{eqnarray}
and
\begin{eqnarray}
S = \frac{e^{i\delta}}{2} \left(
\begin{matrix}
1 & \sqrt{2} & 1 \\
\sqrt{2} & 0 & -{\sqrt{2}} \\
1 & -\sqrt{2} & 1
\end{matrix}
\right)
\end{eqnarray}
where $e^{i\delta} = e^{i(2\alpha_{y,xy}+\alpha_{x,x})}=e^{-0.0274\pi i}$. The the prefactor $e^{i\alpha_{y,xy}}$ in  $T_y$ is chosen such that the statistics for the trivial and fermionic anyons are exact. The phase factor $e^{i\delta}$ in $S$ can be attributed to numerical errors. The quantum dimensions $1, \, \sqrt2, \, 1$ in the $S$ matrix are consistent with the theoretical values of $I,\,\sigma$ and $\psi$.

The chiral central charge can be obtained from $T_y=e^{-{2\pi i\over 24}c_-}{\rm diag}(\theta_1,\theta_\sigma,\theta_\psi)$ or from ${1\over D}\sum_i d_i^2\theta_i = e^{{2\pi i\over 8}c_-}$ (where $D=\sqrt{\sum_id_i^2} =2$). The first relation gives $c_-= -{12\over\pi}\alpha_{y,xy}=1.59$ while the second relation yields $c_-={4\over\pi}(\alpha_{x,x}-\alpha_{y,xy})=1.48$. The inconsistency between these two values can also be attributed to numerical errors. The averaged central charge is $c_-=1.53\pm0.06$.

Above modular matrices as well as the chiral central charge are in excellent agreement with $SO(3)_1$ conformal field theory predictions given in Eqs.~(\ref{ST}) and (\ref{cc}), which verifies that the Gutzwiller-projected weak pairing state is indeed a NACSL of $SO(3)_1$ type.

\subsection{Fractional spin of the non-Abelian anyon}

\begin{figure} [t]
\centering
\includegraphics[width=3.4in]{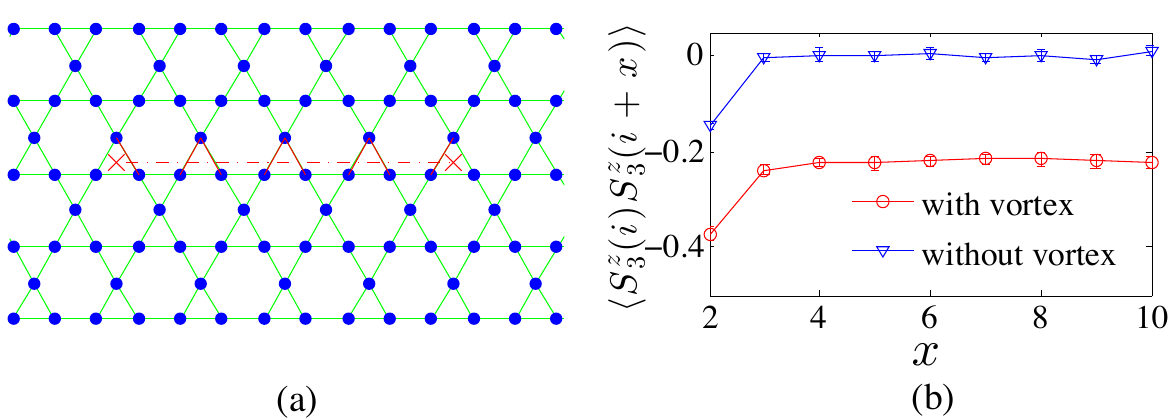}
\caption{(Color online) (a) Two $\pi$-flux vortices are connected with a string. All the mean-field couplings across the string reverse their signs; (b) correlation functions of the `cluster spin' $S^z_3(i)=\sum_{j\in\Delta(i)} S_j^z$, i.e., the total spin of the three vertices of a triangle. The red (blue) line is the case where a vortex is present (absent) at the center of each cluster.}
\label{fig:vortexCorr}
\end{figure}

Although the spin Hall conductance is not fractionalized, the non-Abelian anyons do carry fractional symmetry charge. To see this, we create vortices on top of the $p_x+ip_y$ mean-field state and perform Gutzwiller projections. As shown in Fig.~\ref{fig:vortexCorr}(a), the red dashed line is a string linking two vortices (represented by the crosses). If a bond is crossed by the dashed line, its color changes to red, meaning that the signs of the spinon hopping and pairing are reversed. 
It is known that a vortex (which is essentially a $\pi$-flux for a single fermion) in the $p_x+ip_y$ topological superconductor traps a Majorana zero mode \cite{ReadGreen2000}. There are three types of fermions $c_x, \, c_y, \, c_z$ in our mean-field Hamiltonian, so a vortex will trap three Majorana zero modes $\gamma_x, \, \gamma_y, \, \gamma_z$ ($\{\gamma_m,\gamma_n\}=2\delta_{mn}$) in the weak pairing phase.

The three Majorana operators form an $SO(3)$ vector. For a global spin rotation $e^{i\hat{S}_z\theta}$, the zero modes transform as
\begin{eqnarray}
\label{rotateZ}
e^{-i\mathcal S_z \theta} \left(
\begin{matrix}
\gamma_x \\ \gamma_y \\ \gamma_z
\end{matrix}
\right) e^{i\mathcal S_z \theta} =
\left(\begin{matrix}
\cos\theta & \sin\theta & 0 \\ -\sin\theta & \cos\theta & 0 \\ 0 & 0 & 1
\end{matrix}\right)
\left(\begin{matrix}
\gamma_x \\ \gamma_y \\ \gamma_z
\end{matrix}\right),
\end{eqnarray}
where $\mathcal S_z$ is the operator acting on the degenerate Hilbert space spanned by the Majorana zero modes. It can be checked that the operator
\[
\mathcal S_z=-{i\over2}\gamma_x\gamma_y
\]
satisfies Eq. (\ref{rotateZ}) and one can similarly define $\mathcal S_x=-{i\over2}\gamma_y\gamma_z$ and $\mathcal S_y=-{i\over2}\gamma_z\gamma_x$. This yields
\[
\mathcal S_x^2+\mathcal S_y^2 +\mathcal S_z^2 ={3\over4}={1\over2}\times({1\over2} +1),
\]
so the spin quantum number of the Hilbert space spanned by the three Majorana zero modes is 1/2. 

The two spin-1/2 zero modes in a pair of vortices can form a singlet or a triplet, which are degenerate in energy if the distance between them is infinite. However, if the distance between the vortices is not too much larger than the correlation length (for instance, in our numerical simulation the largest distance between the vortices is 10 unit cells), then the energy of the `zero modes' are exponentially small but not exactly zero. Therefore, the spin-1/2 objects trapped in the vortex cores are weakly coupled, resulting in a exponentially small splitting between the singlet and triplet states (the singlet is slightly lower in energy).

Since the spin quantum number is not affeced by the Gutzwiller projection, it is expected that the non-Abelian anyon corresponding to the vortex carries spin-1/2 angular momentum. To verify this, we define a cluster-spin operator $S^z_{3}$ for the total spin of the three sites on the triangle $\Delta(i)$ where the vortex resides
\[
S^z_3(i)=\sum_{j\in \Delta(i)} S_j^z,
\]
and numerically compute the correlation of two such operators. 

Figure \ref{fig:vortexCorr}(b) shows our results that are computed in the singlet channel on a cylinder with $L_x=20$ and $L_y=4$. If each triangle contains a vortex, then the correlation of the corresponding cluster spin operators converges to a value which is close to $-1/4$ \footnote{The numerical result is not converging to $-{1\over4}$ precisely, because the size of the zero mode trapped by a vortex is usually larger than one triangle, therefore after projection the spin angular momentum of the spin-1/2 zero mode is distributed in an area which is larger than one triangle. So in defining the triangle-cluster-spin operator $S_3^z$, we have dropped the spin angular momentum carried by the tail of the spin-1/2 zero mode.}, whereas it converges to $0$ if there are no vortices in the triangles. This confirms that the non-Abelian anyon carries spin 1/2 \cite{greiter2009}, in analogy to the edge states of $S=1$ Haldane chain \cite{WhiteHuse1993,LiuZhouTuWenNg2012}.

It is important to note that each vortex not only carries a local spin-1/2 degrees of freedom, but also a nonlocal Hilbert space similar to an Ising anyon. In general, braiding of two $\pi$-flux vortices results in a non-local operation (which generates entanglement) together with a local spin rotation for the spin-1/2 degrees of freedom.

Strictly speaking, the NACSL phase is an $SO(3)$ symmetry-enriched topological order\cite{WenSET15}, where the $\sigma$ anyon has two components (spin-up and spin-down) and carries a nontrivial projective representation of $SO(3)$. If a weak magnetic field is applied to break the $SO(3)$ symmetry to $U(1)$ symmetry, which has no nontrivial projective representation, then the local spin-1/2 degeneracy is lift and the $\sigma$ anyon has only one component.

\section{Possible local Hamiltonian and topological phase transition}
\label{sec4}

The non-Abelian Moore-Read Pfaffian state was originally constructed for quantum Hall systems, but it has also been studied in spin-1 systems. In particular, parent Hamiltonian for which the Pfaffian state is the exact ground state have been proposed \cite{CSL2,CSL3}. It contains three-body interactions and long-range interactions. The interesting possibility is that the ground state of a properly truncated Hamiltonian with only short-range interactions falls in the same phase \cite{greiter2009,CSL3}. It is natural to ask whether there exists a local Hamiltonian which realizes the topological order we studied above.

Noticing a natural mean-field decoupling that yields the $p_x+ip_y$-pairing (see Eq.~(\ref{3bd_2}) in Appendix \ref{appsec2}), we consider the model with Hamiltonian
\begin{eqnarray}\label{JKChi}
H = \sum_{\langle ij\rangle} [J_1 \mathbf S_i\cdot \mathbf S_j  - K (\mathbf S_i\cdot \mathbf S_j )^2]+ J_{\chi}\sum_{\vartriangle,\triangledown}(\mathbf S_i\times\mathbf  S_j)\cdot \mathbf S_k,\nonumber\\
\end{eqnarray}
where $J_1,\,K>0$, and $i,\,j,\,k$ goes counterclockwisely on each equilateral triangle (three-body interactions may also exist on skew triangles but are neglected here for simplicity). The $J_\chi$ term preserves $SO(3)$ spin rotation symmetry but explicitly breaks time-reversal symmetry, which may help to stabilize the NACSL (see Appendix \ref{appsec2}).

\begin{figure} [t]
\centering
\includegraphics[width=3.in]{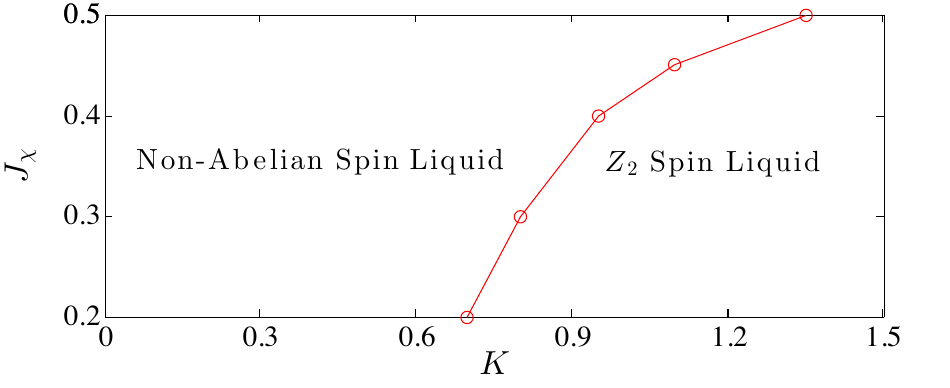}
\caption{(Color online) Tentative phase diagram of the Hamiltonian (\ref{JKChi}). We choose $J_1=1$ and only consider the region where $J_\chi$ is not too small and $K$ is not too large.}
\label{fig:phasedgrm}
\end{figure}

Using the RVB wave functions constructed in Sec.~\ref{sec2} as trial ground states and by minimizing the variational energy with respect to the Hamiltonian in Eq. (\ref{JKChi}), we obtain a tentative phase diagram which contains two spin liquid phases, see Fig.~\ref{fig:phasedgrm}. Part of the numerical results are shown in Tab.\ref{tab:Energy} in Appendix \ref{appGutz}.  In obtaining the phase diagram, an Abelian chiral spin liquid state \cite{LiuMeiYeWen, LauchliCSL15, ShengChen14, DNShengSRe14, DNShengPRB15, Bieri15,JXLiNJP15} (projected Chern band insulator) has also been considered as a trial ground state, but is excluded since its variational energy is generally higher than the NACSL. While the competition between different spin liquid phases is revealed, we cannot rule out the possible existence of symmetry breaking phases \cite{LiSu15, Lauchli15}, such as the spin-nematic phase which may appear when $K$ is sufficiently large. Here we ignore the possible symmetry breaking phases and leave a complete phase diagram for future work. Noticing that the two spin liquid phases in Fig.~\ref{fig:phasedgrm} have the same symmetry but different topological orders, the direct phase transition between these two spin liquids, if indeed takes place, should be a topological one and deserves further study. 

\begin{figure} [t]
\centering
\includegraphics[width=2.in]{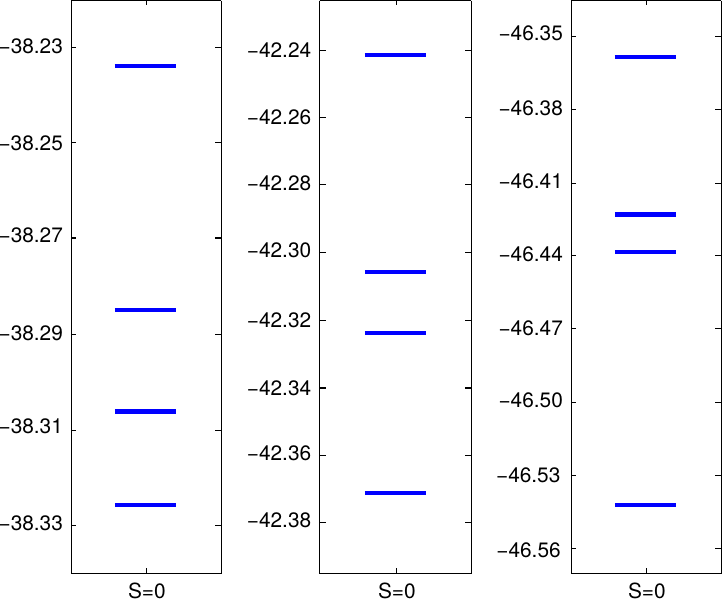}
\caption{Energy spectra on the kagome lattice with $L_x=3$ and $L_y=2$. (a) $J=1$, $K=0.2$, $J_{\chi}=0.3$; (b) $J=1$, $K=0.25$, $J_{\chi}=0.4$; (c) $J=1$, $K=0.3$, $J_{\chi}=0.5$. All the states are spin singlet.}
\label{fig:ExactDiag}
\end{figure}

Although it requires extensive numerical calculations to fully determine the phase diagram of Eq. (\ref{JKChi}), we have made a first step in this direction by performing exact diagonalization on the kagome lattice with $L_x=3$ and $L_y=2$. The energy spectra for three different sets of parameters are given in Fig. \ref{fig:ExactDiag}. It can be argued that there are three quasi-degenerate ground states in certain cases, which is consistent with the theoretical prediction, but the splitting between these states is still obvious. One may hope to see more convincing signatures by studying larger systems using other numerical methods such as tensor network renormalization.

The NACSL has potential applications in topological quantum computation. An important issue is how to localize the non-Abelian anyons. Since the $\pi$-flux vortices trap non-Abelian $\sigma$ anyons, the question becomes how to stabilize the $\pi$-flux vortices using local interactions. The three-body interaction can be written as a ring-exchange term (see Appendix \ref{appsec2}), so the $\pi$-flux vortices can be stabilized by defect triangles with $J_\chi<0$. There may also be three-body terms on skew triangles so anyons can also be localized by defect interactions at such places. As non-Abelian anyons must appear in pairs, we should also create defect three-body interactions in pairs to avoid unexpected degeneracy when trapping the anyons [see Fig.~\ref{fig:vortexCorr}(a)].

\section{Summary and outlook}
\label{sec5}

To summarize, we have studied $S=1$ RVB wave functions  on the kagome lattice by Gutzwiller projecting $p_x+ip_y$ superconductors. By computing various topological quantities, it is demonstrated that a Gutzwiller-projected weak pairing superconductor is a non-Abelian $SO(3)_1$ chiral spin liquid, while the Gutzwiller-projected strong pairing superconductor is an Abelian $Z_2$ spin liquid. A topological quantum phase transition between two different topological orders is observed when continuously tuning the variational parameters in the Gutzwiller wave functions.  We proposed a microscopic model Hamiltonian and provided a preliminary phase diagram by simple VMC calculations. We further performed small-size exact diagonalization of several Hamiltonians in the NACSL phase and nearly-degenerate singlet-ground-states are in support of the VMC phase diagram. We finally proposed that the non-Abelian anyons may be trapped locally using defect three-body interactions.

\section*{Acknowledgement}

We thank Zheng-Cheng Gu, Jia-Wei Mei, Wei Zhu, Meng Cheng, Vic K.~T. Law, Patrick A. Lee and Xiao-Gang Wen for helpful discussions, and acknowledge the computational time in the cluster of IAS in Tsinghua University. This research is supported in part by Perimeter Institute for Theoretical Physics. ZXL thanks the support from NSFC (No.11574392), the Ministry of Science and Technology of China (Grant No. 2016YFA0300504), the Fundamental Research Funds for the Central Universities and the Research Funds of Renmin University of China (No. 15XNLF19). HHT and YHW acknowledge support from the EU Integrated Project SIQS and the DFG through the Excellence Cluster ``Nanosystems Initiative Munich". XJL is supported by the Thousand-Young-Talent Program of China and by NSFC (No.11574008, No.11761161003).  YZ is supported by National Key Research and Development Program of China (No.2016YFA0300202), National Basic Research Program of China (No.2014CB921201), NSFC (No.11774306), the Key Research Program of the Chinese Academy of Sciences (Grant No. XDPB08-4) and the Fundamental Research Funds for the Central Universities in China.
TKN thanks the support from Hong Kong Research Grant Council HKUST3/CRF/13G.

\appendix

\section{Decoupling the spin-spin interactions}
\label{appsec2}

This section is devoted to explain why the mean-field parameters $\chi,\Delta$ can be used as variational parameters. It is known that the two-body spin interactions can be written using fermionic operators as \cite{LiuZhouTuWenNg2012,LiuZhouNgExt}:
\begin{eqnarray}
&& \mathbf S_i\cdot\mathbf S_j = -(\hat \chi_{ij}^\dag \hat \chi_{ij} +\hat \Delta_{ij}^\dag \hat \Delta_{ij}),\nonumber\\
&& (\mathbf S_i\cdot\mathbf S_j)^2 = \hat \Delta_{ij}^\dag \hat \Delta_{ij},
\end{eqnarray}
with $\hat \chi_{ij}=C_i^\dag C_j$ and $\hat\Delta_{ij}=\bar C_i^\dag C_j$. The mean-field decoupling of the $J_1$ and $K$ terms naturally give rise to the mean-field parameters $\chi\sim \langle\hat \chi_{ij}\rangle$ and $\Delta\sim\langle\hat\Delta_{ij}\rangle$.

We next turn to the three-body interaction $({\mathbf S}_i\times{\mathbf S}_j)\cdot{\mathbf S}_k$. In the $c_x, c_y, c_z$ basis, the hopping and pairing operators are
\[
\hat \chi_{ij} = c_{\alpha i}^\dag c_{\alpha j},\ \hat \Delta_{ij} = -c_{\alpha i} c_{\alpha j},\
\]
and the spin operators are
\[
S^\alpha=-i\varepsilon^{\alpha\beta\gamma}c_\beta^\dag c_\gamma,
\]
where $\alpha,\beta,\gamma=x,y,z$ and repeated indices are summed (the same convention is used below). The three-body interaction can be written as
\begin{eqnarray}\label{3bd}
(\mathbf S_i\times\mathbf  S_j)\cdot \mathbf S_k &=& \varepsilon^{\alpha\beta\gamma}S_i^\alpha S_j^\beta S_k^\gamma\nonumber\\
&=& i \varepsilon^{\alpha\beta\gamma}\varepsilon^{\alpha\mu\nu}\varepsilon^{\beta\rho\eta}\varepsilon^{\gamma\lambda\sigma}c_{\mu i}^\dag c_{\nu i} c_{\rho j}^\dag c_{\eta j} c_{\lambda k}^\dag c_{\sigma k}.\nonumber\\
\end{eqnarray}
and further simplified to
\begin{eqnarray}\label{3bd_2}
(\mathbf S_i\times\mathbf  S_j)\cdot \mathbf S_k &=&i\left[\left(\hat \chi_{ij}\hat\Delta_{jk}^\dag\hat\Delta_{ki} +
\hat\Delta_{ij}\hat\chi_{jk}\hat\Delta_{ki}^\dag \right.\right.\nonumber\\ &&\left.\left.+ \hat\Delta_{ij}^\dag\hat\Delta_{jk}\hat\chi_{ki}
- \hat \chi_{ij}\hat \chi_{jk}\hat \chi_{ki}\right)-{\rm h.c.}\right]\nonumber\\
\end{eqnarray}
using the relations
\[
\varepsilon^{\alpha\beta\gamma}\varepsilon^{\alpha\mu\nu}=\delta_{\beta\mu}\delta_{\gamma\nu}-\delta_{\beta\nu}\delta_{\gamma\mu}
\]
and
\begin{eqnarray*}
\varepsilon^{\beta\rho\eta}\varepsilon^{\gamma\lambda\sigma}&=&\delta_{\beta\gamma}(\delta_{\rho\lambda}\delta_{\eta\sigma}-\delta_{\rho\sigma}\delta_{\eta\lambda})\\
&& -\delta_{\beta\lambda}(\delta_{\rho\gamma}\delta_{\eta\sigma}-\delta_{\rho\sigma}\delta_{\eta\gamma})\\
&& +\delta_{\beta\sigma}(\delta_{\rho\gamma}\delta_{\eta\lambda}-\delta_{\rho\lambda}\delta_{\eta\gamma})
\end{eqnarray*}
This means that the $J_\chi$ term can be decoupled using the two parameters $\chi$ and $\Delta$.

One can see from Eq. (\ref{3bd_2}) that, if there is a vortex in the triangle $(ijk)$, the values of $\langle \hat \chi_{ij}\rangle$ and $\langle\hat \Delta_{ij}\rangle$ on one of the three bonds (i.e., $ij$, $jk$ and $ki$) reverse its sign. As a result, the value of $\langle J_\chi(\mathbf S_i\times\mathbf  S_j)\cdot \mathbf S_k\rangle$ on the triangle also reverses its sign, so the state has a higher energy compared to the ground state. If we reverse the sign of $J_\chi$ on the triangles which contain $\pi$-flux vortices, $\langle -J_\chi(\mathbf S_i\times\mathbf  S_j)\cdot \mathbf S_k\rangle$ will have a low energy and the state with anyons localized in the vortices becomes the ground state of the new Hamiltonian.

\section{Gutzwiller Projection of BCS states}\label{appGutz}

The ground state of the mean-field Hamiltonian (\ref{MF}) is a BCS wave function
\[
|{\rm BCS}\rangle = \prod_{i>j}\left[1+a_{ij}(c_{1i}^\dag c_{-1j}^\dag - c_{0i}^\dag c_{0j}^\dag +c_{-1i}^\dag c_{1j}^\dag)\right] |{\mathrm {vac}}\rangle,
\]
where $a_{ij}=-a_{ji}$ is the 
paring amplitude of two spinons in a ``Cooper pair" which decays exponentially with their relative distance in the strong pairing case and decays in power law $a_{ij}\propto|\mathbf r_i-\mathbf r_j|^{-1}$ in the weak paring case. 

A practical way of constructing a spin liquid wave function from above mean-field state is to enforce the particle number constraint through Gutzwiller projection
\begin{eqnarray}
|{\rm RVB}\rangle &=& P_G |{\rm BCS}\rangle\nonumber\\
&=& \sum_{\alpha} \mathrm{det}[A^{(1,-1)}] \mathrm{Pf}[B^{(0,0)}] |\alpha\rangle,
\end{eqnarray}
where $|\alpha\rangle$ is an Ising configuration, $P_G$ is the Gutzwiller projection operator that enforces the particle number constraint. 
The matrices $A^{(1,-1)}$ and $B^{(0,0)}$ are
\begin{eqnarray*}
A^{(1,-1)} = \left(\begin{matrix}
a_{m_1n_1} & a_{m_1n_2} & ... \\
a_{m_2n_1} & a_{m_2n_2} & ... \\
\vdots & \vdots & \ddots      \\
\end{matrix}\right),\\
B^{(0,0)} = \left(\begin{matrix}
0 & -a_{p_1p_2} & ... \\
-a_{p_2p_1} & 0 & ... \\
\vdots & \vdots & \ddots \\
\end{matrix}\right),
\end{eqnarray*}
where $m_i, \, n_i, \, p_i$ are the positions of $c_1, \, c_{-1}, \, c_0$ fermions in the configuration $|\alpha\rangle$, respectively. The particle number of $c_1$ is equal to that of $c_{-1}$ to ensure that the ground state has total spin $S_z=0$.

In the verctor bases $c_x, c_y, c_z$ introduced in Eq.~(\ref{vector}), the mean-field ground state becomes
\begin{eqnarray*}
|{\rm BCS}\rangle &=& \prod_{i>j}\left[1-a_{ij}(c_{xi}^\dag c_{xj}^\dag + c_{yi}^\dag c_{yj}^\dag +c_{zi}^\dag c_{zj}^\dag)\right] |{\mathrm {vac}}\rangle\\
&=& \prod_{r>s}(1-a_{rs}c_{xr}^\dag c_{xs}^\dag) \prod_{u>v}(1-a_{uv}c_{yu}^\dag c_{yv}^\dag) \\
&\times& \prod_{p>q}(1 - a_{pq}c_{zp}^\dag c_{zq}^\dag) |{\mathrm {vac}}\rangle
\end{eqnarray*}
which is essentially three copies of $p_x+ip_y$ superconductors. The projected state can be written as
\begin{eqnarray*}
|{\rm RVB}\rangle &=& P_G |{\rm BCS}\rangle\nonumber\\
&=& \sum_{\alpha'} \mathrm{Pf}[C^{(x,x)}]\mathrm{Pf}[D^{(y,y)}] \mathrm{Pf}[B^{(z,z)}]|\alpha'\rangle \nonumber
\end{eqnarray*}
where $\alpha'$ is the spin configuration created by $c_x^\dag, \, c_y^\dag, \, c_z^\dag$. The matrices $B,\,C,\,D$ are defined as $$B^{(z,z)}=B^{(0,0)}$$ and
\begin{eqnarray*}
C^{(x,x)} &=& \left(\begin{matrix}
0 & -a_{m_1m_2} & ... \\
-a_{m_2m_1} & 0 & ... \\
\vdots & \vdots & \ddots \\
\end{matrix}\right),\\
D^{(y,y)} &=& \left(\begin{matrix}
0 & -a_{n_1n_2} & ... \\
-a_{n_2n_1} & 0 & ... \\
\vdots & \vdots & \ddots \\
\end{matrix}\right),
\end{eqnarray*}
where $m_i,\,n_i,\,p_i$ are the positions of the $c_x,\,c_y,\,c_z$ fermions in the configuration $|\alpha'\rangle$, respectively. The sum of the particle numbers of $c_x, c_y, c_z$ fermions is equal to the number of lattice sites.

Using the RVB states with two variational parameters $\Delta, \lambda$ (the parameter $\chi$ is normalized as $\chi=1$) as trial wave functions and by minimizing the expectation energy of the Hamiltonian in eq. (\ref{JKChi}), we can obtain the optimized variational parameters with different interaction strengths. From these information we obtain a preliminary phase diagram (see Fig.\ref{fig:phasedgrm}). Part of the numerical results are listed in table \ref{tab:Energy}. It should be mentioned that the values of the optimized `ground state' energies in table \ref{tab:Energy} are only of significance within the variational approach when different trial states are compared. These energies might be away from the true ground state energies since the Gutzwiller projected wave functions only contain two variational parameters and this variational approach is based on mean-field approximations where the fluctuations of mean-field parameters are ignored.

\begin{table}[t]
\centering
\begin{tabular}{c|c||c|c|c}
\hline
$\ \ K\ \ $ & $\ \ J_{\chi}\ \ $ &\ \ \ Energy \ \ \ & \ \ \ \ \ \ $\Delta$\ \ \ \ \ \ &\ \ \ \ \ \ $\ \ \lambda$\ \ \ \ \ \  \\
\hline
0     & 0            & -1.1880  & 0.8467   & 0.4155 \\
\hline
0     & 0.2         & -1.2829  & 0.9319  & 0.3271 \\
0.2  & 0.2         & -1.9575  & 1.5102   &-0.0712 \\
0.8  &  0.2        & -4.4343  & 50.0000 & 4.9456 \\
\hline
0     &  0.3        & -1.3311  & 0.9530   & 0.2349 \\
0.2  &  0.3        & -2.0038  & 1.3998   &-0.0462 \\
1     &  0.3        & -5.2961  & 69.8872 & 4.4248 \\
\hline
0     &  0.5        &-1.4283   &0.9863    & 0.2659 \\
0.3  &  0.5        & -2.4515  &1.5622    & -0.0330\\
1     &  0.5        & -5.2987  &22.7579  & 1.7416 \\
1.5  &  0.5        &-7.4513   &90.2659  & 4.1193 \\
\hline
\end{tabular}

\caption{Variational ground state energies and variational parameters of the model (\ref{JKChi}) with different interaction parameters, where the Heisenberg interaction strength $J$ is normalized $J=1$. The parameter $\chi$ is also normalized $\chi=1$ and there are only two independent variational parameters $\Delta, \lambda$ remaining. The parameter $\lambda<4$ means a non-Abelian Chiral spin liquid state, while $\lambda>4$ stands for a $Z_2$ spin liquid state. } \label{tab:Energy}
\end{table}

\section{Effective Chern-Simons field theory}
\label{appCS}

Due to interactions between the fermions, the mean-field parameters $\chi$, $\Delta$, $\lambda$ are subject to fluctuations. The amplitude fluctuations are gapped at low energy, so we only need to consider the phase fluctuations, which behave like gauge fields coupling to the fermionic spinons. The pairing of fermions Higgses the $U(1)$ gauge fluctuations and only $Z_2$ gauge symmetry remains. This gives the following Hamiltonian with $Z_2$ gauge fluctuations
\begin{eqnarray}\label{MFZ2gauge}
H_{\rm mf}(\sigma_{ij},\delta\lambda_i)&=& \sum_{\langle ij \rangle} \sigma_{ij}[\chi C_i^\dag C_j + \Delta e^{\theta_{ij}}C_i^\dag \bar C_j + {\rm h.c.}] \nonumber\\
&&+ \sum_i(\lambda +\delta\lambda_i)(C_i^\dag C_i-1),
\end{eqnarray}
where $\sigma_{ij}$ are spatial components of the $Z_2$ gauge fields and the temporal fluctuations $\delta\lambda_i$ is continuous. 

If the superconductors have vanishing Chern numbers, i.e. in the strong pairing case,  then the spinon fields can be integrated out straightforwardly, giving rise to a {\it $Z_2$ gauge theory} as the low-energy effective field theory of the system.

On the other hand, if the superconductors have nontrivial Chern numbers, i.e. in the weak pairing case,  a Chern-Simons term is expected after integrating out the spinon fields. Before going to the intrinsic field theoretical description, we firstly investigate  the response theory of the system. Owing to the nontrivial Chern numbers, the system would exhibit a nontrivial Hall effect when it is probed by symmetry twisting fields.

As seen from the mean-field theory, the spinons will couple to not only the gauge fluctuations but also the symmetry twisting fields (behaving like $SO(3)$ gauge fields)
\begin{eqnarray}\label{MFprob}
H_{\rm mf}(\sigma_{ij},A_{ij})&=& \sum_{\{i,j\}} \sigma_{ij}[\chi_{ij} C_i^\dag e^{iA_{ij}}C_j + \Delta_{ij}C_i^\dag e^{iA_{ij}} \bar C_j \nonumber\\
&&+ {\rm h.c.}] + \sum_i(\lambda_i +\delta\lambda)(C_i^\dag C_i-1),
\end{eqnarray}
where $A_{ij}=A^x_{ij}S^x+A^y_{ij}S^y+A^z_{ij}S^z$ is the external $SO(3)$ probing field related to spin rotation symmetry. After integrating out the internal gauge fluctuations and the fermions, we expect to get the following $SO(3)_1$ Chern-Simons field theory as the response theory (in the imaginary time formalism and in the continuum limit)
\begin{eqnarray}\label{prob}
\mathcal L_{\rm res} = i{k\over 4\pi} {1\over2}{\rm Tr}[\varepsilon^{\mu\nu\lambda}(A_\mu\partial_\nu A_\lambda) - {1\over 3}A^3] + {\cal L}_{\rm Maxwell} + ... ,\nonumber\\
\end{eqnarray}
where $k=1$ is equal to the Chern number of each species of fermions $c_x,c_y,c_z$. If the probing field only has a $z$-component such that $A_\mu=A^z_\mu S^z$, the spin Hall current is
$
J^z_\mu={\delta \mathcal L_{\rm res}\over \delta A^z_\mu} = {1\over 2\pi} F^z_{\mu},
$
where $F^z_\mu=i\varepsilon^{\mu\nu\lambda}(\partial_\nu A^z_\lambda)$ is the probing field strength. This gives a spin Hall conductance 1 in units of ${1\over 2\pi}$. For comparison, we point out that the spin Hall conductances of bosonic $U(1)$ symmetry-protected topological phases and $S=1$ Abelian chiral spin liquids are even integers in units of ${1\over 2\pi}$ \cite{ChenGuLiuWen2011, LuVishwanath2012, LiuWen2012, LiuMeiYeWen}.

Although above field theory is a response theory, it can be argued that the intrinsic field theoretical description of the system (in the weak pairing case) is still an $SO(3)_1$ Chern-Simons theory. The reason is the following. The action (\ref{prob}) is not gauge invariant if it is defined on an open manifold with a boundary. The gauge anomaly can be canceled by anomalous matter fields, namely three species of chiral Majorana fermions (see, for example, Ref.~\onlinecite{LiuWen2012}) moving along the boundary, which carries chiral central charge $c_-={3\over2}$ and can be described by a conformal field theory --- the $SO(3)_1$ chiral Wess-Zumino-Witten theory. The intrinsic field theory, if exist, can also describe this anomalous boundary modes. A natural answer is still the {\it $SO(3)_1$ Chern-Simons gauge theory},  by replacing the $SO(3)$ symmetry-twisting fields $A_{\mu}$ in (\ref{prob}) with the intrinsic $SO(3)$ gauge fields $a_{\mu}$ (where the strength of $a_\mu$ describes the spin current) 
\begin{eqnarray}\label{intrin}
\mathcal L_{\rm eff} = i{1\over 4\pi} {1\over2}{\rm Tr}[\varepsilon^{\mu\nu\lambda}(a_\mu\partial_\nu a_\lambda) - {1\over 3}a^3] + {\cal L}_{\rm Maxwell} + ... ,\nonumber\\
\end{eqnarray}
This bulk Chern-Simons Lagrangian together with the boundary chiral Majorana modes are $SO(3)$ gauge invariant and provide a complete intrinsic field theoretical description of the  whole system.

The above $SO(3)_1$ Chern-Simons field theory contains Ising-like non-Abelian anyons as its intrinsic elementary excitations, which indicates that in the weak pairing case the spin system belongs to a non-Abelian chiral spin liquid phase.

\bibliography{NACSL}

\end{document}